\documentclass{aa}
\usepackage{amsmath,graphics,amssymb}
\usepackage[varg]{txfonts}
\usepackage{natbib}
\usepackage{times}
\usepackage[hidelinks]{hyperref}

\newfont{\eufont}{eufm10}

\newcommand{\zav}[1]{\left(#1\right)}

\newcommand{\kms}{\ensuremath{\text{km}\,\text{s}^{-1}}}
\newcommand{\vel}{{v}}

\newcommand{\ms}{\ensuremath{{M}_{\odot}}}
\newcommand{\msr}{\ensuremath{\ms\,\text{yr}^{-1}}}
\newcommand\vnek{\ensuremath{\varv_\infty}}

\newcommand{\teff}{\ensuremath{T_\text{eff}}}

\newcommand{\ch}{\text{cl}}

\newcommand\cc{\ensuremath{C_\text{c}}}

\begin{document}

\title{The origin of dispersion in the wind mass-loss rate--luminosity relation of massive
O stars}
\titlerunning{Dispersion in the wind mass-loss rate--luminosity relation}

\author{J.~Krti\v{c}ka \and I.~Krti\v ckov\'a}

\institute{Department of Theoretical Physics and Astrophysics, Faculty of
           Science, Masaryk University, CZ-611 37 Brno,
           Czech Republic}

\date{Received}

\abstract{ Stellar winds of hot stars are accelerated by the radiative force
that mainly comes from the line transitions of heavy elements. As a result, the
wind mass-loss rate is expected to scale mostly with the stellar luminosity.
This scaling clearly emerges from the observational analysis of O stars and
their winds, but with substantial scatter. The origin of this scatter is
unclear.} {Part of the scatter of the observational relation between the hot
star wind mass-loss rate and luminosity may originate from the sensitivity of
the mass-loss rate to other stellar parameters besides the luminosity, such as
the effective temperature, mass, and metallicity. We tested this possibility
using our wind models.} {We calculated tailored wind models for stellar
parameters corresponding to selected individual O stars from the Large
Magellanic Cloud as derived from a previous analysis. To model the wind, we used
our global wind code that calculates the radiative force from the solution of
the radiative transfer equation in the comoving frame.} {Wind models calculated
for particular stellar parameters reproduce the observational mass-loss rates
better than simple scaling relations derived from model grids. This indicates
that the dependence of the mass-loss rate on the other stellar parameters except
the luminosity might explain part of the dispersion in the relation between the
mass-loss rate and luminosity. However, some dispersion still remains, which
might indicate that wind variability is another source of the dispersion.}{}

    \keywords{stars: winds, outflows --
              stars:   mass-loss  --
              stars:  early-type --
              hydrodynamics --
              supergiants --
              Magellanic Cloud
}

\maketitle
\nolinenumbers

\section{Introduction}

The stellar wind of hot stars is accelerated by the radiative force originating in
the absorption and scattering of stellar radiation either due to line
transitions or Thomson scattering \citep{lusol,cak}. As a result, the wind
mass-loss rate, which gives the amount of mass lost per unit of time, mostly depends
on the stellar luminosity. This relation clearly emerges from the
observational studies of O star winds \citep{hezkysedi,hawvmm}, which show a
relatively tight correlation between the wind mass-loss rates and the luminosities of
individual stars derived from observational analysis. However, the same studies
clearly point to a considerable scatter in the relation between mass-loss
rates and luminosities that can amount to a factor of a few. The origin of this
scatter remains unclear.

One of the sources of the scatter in the observational wind mass-loss
rate--luminosity relation could be time variability of the wind line
profiles, which are used as observational probes of the mass-loss rate
\citep{masvar,parvar}. As a result of this variability, the observed wind line
profile is just a single realization of possible line profiles from a whole
ensemble, likely leading to the scatter in the observational mass-loss rate
estimates. Another possible source of the scatter is the process of the
mass-loss rate determination itself. The hot star wind mass-loss rates are not
derived from some simple observational indicators. Instead, sophisticated codes
are used to simulate the wind spectra and to reproduce the observed wind line
profiles by varying the wind parameters, one of which is the mass-loss rate
\citep[e.g.,][]{sanporov}. The scatter might in particular be connected with the
treatment of small-scale inhomogeneities (clumping) in the spectrum synthesis
codes \citep{clres1,supujo}. The free parameters used to describe clumping show
large scatter for individual stars, and even an improved treatment of the clumping
that accounts for optically thick clumps does not lead to a systematic shift in
the wind mass-loss rates \citep{preduli,brandxii,zcmfkont}.

The fact that the wind mass-loss rate does not depend just on the stellar
luminosity has been well known since the dawn of the modern stellar wind theory
\citep{cak}. This is reflected in prescriptions of the wind mass-loss rate
\citep{vikolamet,bjorevol,receptpauli} that introduced a dependence of the
wind-mass loss rate on other stellar parameters such as metallicity, effective
temperature, and mass. Even such multiparameter fits to theoretical
predictions show reasonable scatter, however, whose origin remains unclear
\citep[e.g.,][]{vikolamet,bjorevol}. The scatter becomes particularly strong at
low metallicities $Z<0.1Z_\odot$ \citep{zcmfkont}.

The scatter in observational mass-loss rate -- luminosity relation and the
inability of typically polynomial fits to precisely fit the mass-loss rate
predictions may have the same origin. The absence of simple relations even in
theoretical computations indicates that the scatter might be inherent. The
radiative force due to line transitions depends on the ionization and excitation
structure of the wind and on the radiative field. Therefore, the radiative force
in a given location in the wind might depend on the stellar parameters in a very
complicated manner. The best-known example of this sensitivity is the so-called
bistability jump in wind mass-loss rates and terminal velocities, which is
predicted to appear at about 20\,kK due to recombination of \ion{Fe}{iv} to
\ion{Fe}{iii} \citep{bista,vinbisja,bcmfkont}. Similar but weaker variations
can also appear due to other elements.

The existence of the scatter in mass-loss rate -- luminosity relation, or
even more importantly, in parametric fits of the theoretical estimates, is
troublesome. These theoretical mass-loss rate recipes are used in evolutionary
calculations \citep[e.g.][]{alexcentral} and for our understanding of the
interaction between stars and interstellar environment \citep{kobul}.
The uncertainties in the mass-loss rate estimates spread to evolutionary predictions
\citep{kostel,renzovit,agrrozdil,josylva,atlasvetrui}. Therefore, our
understanding of the possible scatter in the mass-loss rate prescription is important
for different fields of astrophysics.

Our study focuses on stars whose stellar and wind parameters were determined by
\citet{hawvmm} from the combination of optical and ultraviolet spectroscopy. We
compare wind mass-loss rate predictions derived for actual stellar parameters
with results derived from a simplified formula that fits results of the same
code on a grid of stellar parameters. In turn, this will help us to understand
whether the scatter in mass-loss rate -- luminosity relation is real.

\section{Comparison with the mass-loss rate formula}

\begin{figure}
\includegraphics[width=0.5\textwidth]{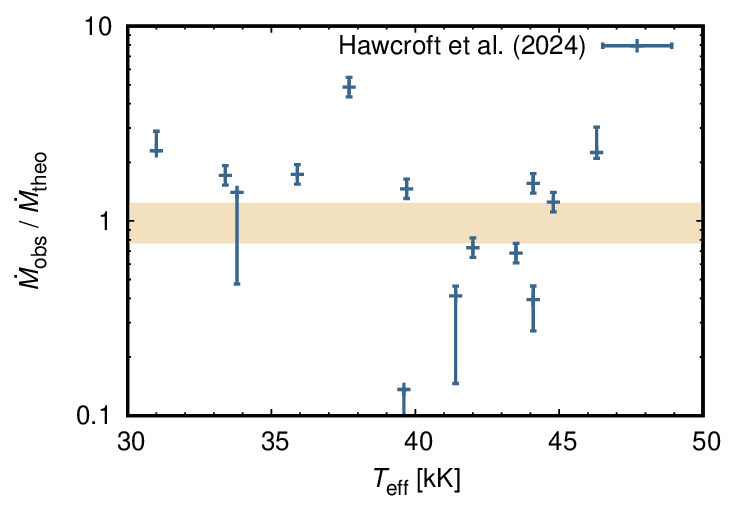}
\caption{Ratio of the mass-loss rates determined from observations
\citep{hawvmm} and from the fitting formula \citep{zcmfkont} evaluated for estimated
stellar parameters, plotted as a function of the stellar effective temperature. The
shaded region denotes the uncertainty of the theoretical estimates implied by
the uncertainties of the observational parameters.}
\label{hawfit}
\end{figure}

Theoretical mass-loss rate predictions of luminous OB stars reproduce the
observational results reasonably well, but with significant scatter. This is
demonstrated in Fig.~\ref{hawfit}, which plots the ratio of mass-loss rates
determined from an ultraviolet and optical wind line analysis for selected O stars
from the Large Magellanic Cloud (LMC) by \citet{hawvmm} and their theoretical
estimates based on the fitting formula of \citet[Eq.~3]{zcmfkont}. The fitting
formula is based on the results of the global wind code with radiative force
from the solution of the radiative transfer equation in the comoving frame
\citep{cmfkont}, and it provides an estimate of the mass-loss rate as a function
of stellar luminosity, effective temperature, and metallicity. We restricted the
\citeauthor{hawvmm} sample to the stars that do not display the so-called
weak-wind problem \citep{bourak,martin,linymarko,mirslabweb}.

For our comparison, we assumed an LMC relative abundance of elements heavier than
helium of 0.5 with respect to the solar value $Z_\odot$
\citep[e.g.,][]{kopec,ven,rolles,bourak}. The fitting formula of
\citet{zcmfkont} underestimates the observational mass-loss rates from
\citet{hawvmm} by a factor of $1.49$, but with a significant root mean square
(rms) of the residuals of $1.18$. A similar rms of the residuals follows from independent
mass-loss rate predictions of \citet{vikolamet} and \citet{bjorevol}.

Mass-loss rate predictions are always expressed as functions of stellar
parameters determined from observations. However, these parameters are subject
to uncertainties related to their measurement. This implies that a comparison with
observations should account for the uncertainties of the mass-loss rate predictions
that stem from the uncertainties in the determination of the basic stellar parameters
that enter these predictions. The related uncertainty of the theoretical estimates
is plotted in Fig.~\ref{hawfit} as the shaded region. The uncertainty was
estimated using the mean error of individual parameters from \citet{hawvmm}.
Apparently, this uncertainty is significantly lower than the dispersion of
observational values, and it cannot by itself explain the dispersion in the ratio of
the observational and theoretical estimates.

\section{Comparison with tailored models}

Theoretical wind codes are typically not directly used for the comparison
with observations. Instead, the codes are executed to calculate a grid of wind
models for specified wind parameters. The outputs of these models (most
frequently, the mass-loss rates) are then fitted as functions of stellar
parameters, and these fits are further used for subsequent applications
\citep[e.g.,][]{vikolamet,bjorevol,zcmfkont}. However, the fitting typically
involves a substantial scatter, which indicates that the fitting oversimplifies
the outputs of the models. To test this, we calculated wind models with
parameters that correspond to observational estimates of LMC O-star parameters
from \citet{hawvmm}.

\begin{table*}
\caption{Parameters of individual stars from \citet{hawvmm} supplemented
with the wind parameters estimated from tailored models.}
\centering
\label{hvezpar}
\begin{tabular}{rcccccccccccc}
\hline
\hline
ID & Spectral & \multicolumn{1}{c}{$\teff$} & \multicolumn{1}{c}{$R_{*}$} & $M$ &
$\epsilon_\text{He}$ & $\epsilon_\text{C}$ & $\epsilon_\text{N}$ &
$\epsilon_\text{O}$ & $\dot M_{0.5Z_\odot}$ & $\varv_{\infty, 0.5Z_\odot}$ &
$\dot M$ & \vnek \\
&type & \multicolumn{1}{c}{$[\text{K}]$} & \multicolumn{1}{c}{$[{R}_{\odot}]$} &
$[{M}_{\odot}]$ & & & & & [\msr] & [\kms] & [\msr] & [\kms] \\
\hline
087 & O9.7Ib-II       & 31000 & 14.3 & 16.7 & $-1.0$ & $-4.5$ & $-4.2$ & $-4.5$ & $1.1\times10^{-7}$ & 1340 & $9.7\times10^{-8}$ & 1390 \\
440 & O6-6.5II(f)     & 33400 & 19.7 & 19.1 & $-0.9$ & $-4.1$ & $-3.8$ & $-4.4$ & $1.2\times10^{-6}$ &  500 & $7.7\times10^{-7}$ &  690 \\
223 & O9.5IV          & 33800 &  9.1 & 21.9 & $-1.0$ & $-4.2$ & $-4.5$ & $-5.7$ & $1.8\times10^{-8}$ & 1200 & $1.5\times10^{-8}$ &  930 \\
664 & O7II(f)         & 35900 & 15.3 & 24.6 & $-1.1$ & $-4.0$ & $-4.2$ & $-4.1$ & $4.2\times10^{-7}$ & 1020 & $3.9\times10^{-7}$ & 1120 \\
385 & O4-5V           & 37700 & 10.9 & 16.8 & $-1.2$ & $-4.3$ & $-4.8$ & $-3.1$ & $2.7\times10^{-7}$ & 1120 & $2.8\times10^{-7}$ & 1090 \\
184 & O6.5Vnz         & 39600 &  7.1 & 20.1 & $-1.0$ & $-4.1$ & $-5.5$ & $-5.7$ & $5.5\times10^{-8}$ & 2260 & $2.4\times10^{-8}$ & 2180 \\
244 & O5III(n)(fc)    & 39700 & 13.0 & 23.4 & $-1.0$ & $-4.7$ & $-4.5$ & $-4.1$ & $6.9\times10^{-7}$ & 1050 & $5.2\times10^{-7}$ & 1090 \\
096 & O6V((n))((fc))z & 41400 & 16.0 & 83.2 & $-1.0$ & $-4.9$ & $-4.8$ & $-4.9$ & $3.9\times10^{-7}$ & 2630 & $3.1\times10^{-7}$ & 2620 \\
422 & O4III(f)        & 42000 & 13.1 & 20.2 & $-1.0$ & $-3.3$ & $-3.8$ & $-5.0$ & $1.5\times10^{-6}$ &  740 & $1.3\times10^{-6}$ &  680 \\
608 & O4III(f)        & 43500 & 14.8 & 59.2 & $-0.9$ & $-4.7$ & $-3.7$ & $-4.5$ & $5.9\times10^{-7}$ & 2210 & $6.3\times10^{-7}$ & 1810 \\
180 & O3If*           & 44100 & 16.5 & 41.4 & $-0.5$ & $-4.6$ & $-3.1$ & $-6.0$ & $2.4\times10^{-6}$ &  900 & $1.7\times10^{-6}$ & 1090 \\
216 & O4V((fc))       & 44100 & 13.4 & 36.0 & $-1.0$ & $-4.6$ & $-4.7$ & $-3.7$ & $1.0\times10^{-6}$ & 1090 & $8.0\times10^{-7}$ & 1410 \\
143 & O3.5V((fc))     & 44800 &  9.5 & 26.1 & $-1.0$ & $-4.1$ & $-5.5$ & $-4.3$ & $3.4\times10^{-7}$ & 1570 & $3.3\times10^{-7}$ & 1680 \\
586 & O4V((n))((fc))z & 46300 &  7.9 & 18.9 & $-1.0$ & $-4.4$ & $-5.6$ & $-3.1$ & $3.3\times10^{-7}$ & 1530 & $3.0\times10^{-7}$ & 1450 \\
\hline
517 & O9.5V-III((n))  & 31300 & 10.6 & 17.1 & $-1.0$ & $-3.9$ & $-4.4$ & $-4.5$ & $3.4\times10^{-8}$ & 1660 & $3.4\times10^{-8}$ & 1550 \\
235 & O9.7III         & 31500 &  6.5 & 19.8 & $-1.0$ & $-3.9$ & $-5.0$ & $-4.7$ & $2.7\times10^{-9}$ & 2720 & $9.3\times10^{-10}$& 3810 \\
627 & O9.7V           & 32000 &  6.5 & 10.9 & $-0.9$ & $-3.7$ & $-4.1$ & $-5.9$ & $7.6\times10^{-9}$ & 1710 & $8.0\times10^{-9}$ & 1510 \\
280 & O9V             & 33400 &  7.7 & 11.9 & $-1.0$ & $-3.7$ & $-5.3$ & $-3.3$ & $1.8\times10^{-8}$ & 1160 & $1.4\times10^{-8}$ & 1570 \\
\hline
\end{tabular}
\tablefoot{The horizontal line separates stars with normal mass-loss rates from
stars that display the weak-wind problem.}
\end{table*}

We used our global wind code METUJE \citep{cmfkont} to derive wind models for
stars from the \citet{hawvmm} sample. The code METUJE solves hydrodynamical
equations of continuity and momentum and the energy equation from the hydrostatic
photosphere to the supersonic wind assuming spherical symmetry and stationarity. The
level occupation numbers were determined from the kinetic (statistical)
equilibrium equations. The corresponding model atoms were typically adopted from
the code TLUSTY \citep{ostar2003,bstar2006}. The solution of the radiative
transfer equation, which determines the radiative force and radiative heating
and cooling terms \citep{kpp}, was performed in the comoving frame \citep{mikuh}. The code METUJE solves all equations iteratively by relying on the
stellar atmosphere code TLUSTY \citep{tlusty0} for providing the initial
estimate of atmospheric structure. The models were parameterized in terms of the
stellar effective temperature $\teff$, mass $M$, radius $R_*$, and elemental
abundances.

We calculated tailored wind models with all stellar parameters set to values
determined by \citet{hawvmm}, except for the metallicity. To understand the effect
of star-to-star metallicity variations, we either adopted He, C, N, and O
abundances from \citet[obtaining the mass-loss rate $\dot M$ and terminal
velocity $\varv_\infty$]{hawvmm}  or we used the scaled LMC composition of
$0.5Z_\odot$ for CNO elements and the solar value for He (giving the mass-loss rate
and terminal velocity $\dot M_{0.5Z_\odot}$ and $\varv_{\infty, 0.5Z_\odot}$,
respectively). In both cases, the abundances of the remaining elements
were fixed to 0.5 of the solar value \citep{asp09}.

\begin{table}[t]
\caption{Mean ratio of the observational and theoretical mass-loss rates $r$ and
the rms of the residuals $\chi$ (Eq.~\ref{peknarovnice}).}
\label{peknatabulka}
\centering
\begin{tabular}{lcccccc}
\hline
Prediction & $r$ & $\chi$ \\
\hline
Fitting formula of \citet[Eq.~(3)]{zcmfkont} & 1.49 & 1.2 \\
Tailored models with metallicity $0.5Z_\odot$ & 0.86 & 0.7 \\
Tailored models with actual metallicity & 0.99 & 0.7 \\
\hline
\end{tabular}
\end{table}

\begin{figure}
\includegraphics[width=0.5\textwidth]{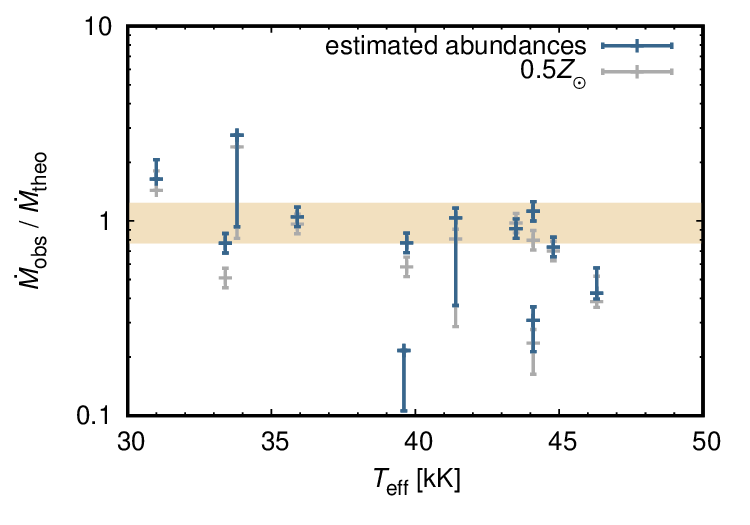}
\caption{Same as Fig.~\ref{hawfit} but for the ratio of the mass-loss rates
determined from observations \citep{hawvmm} and tailored METUJE models computed
for estimated stellar parameters. The models assumed He, C, N, and O
abundances either estimated from observations or scaled by the mean LMC
metallicity (keeping $0.5Z_\odot$ for the rest of the elements).}
\label{hawmet}
\end{figure}

The resulting predicted wind parameter are listed in Table~\ref{hvezpar}
together with the adopted stellar parameters. The elemental abundances are given as
relative number densities with respect to hydrogen,
$\epsilon_\text{el}=\log(n_\text{el}/n_\text{H})$. To compare the results, we
additionally calculated the mean ratio of the observational mass-loss rate
determinations and theoretical predictions and the rms of this ratio,
\begin{align}
r&=\langle\dot M_\text{obs}/\dot M_\text{theo}\rangle,\\\label{peknarovnice}
\chi&=\sqrt{\langle(\dot M_\text{obs}/\dot M_\text{theo}-r)^2\rangle}.
\end{align}
These values are given in Table~\ref{peknatabulka} and are further illustrated in
Fig.~\ref{hawmet}.

It follows from Figs.~\ref{hawfit}, \ref{hawmet}, and Table~\ref{peknatabulka} that
models calculated with observationally determined stellar parameters agree better with
the mass-loss rate estimated from observations than predictions based on
the fitting formula. While basically no observational mass-loss rate appears in the
range of theoretical uncertainties in Fig.~\ref{hawfit}, nearly the opposite is
true in Fig.~\ref{hawmet}, where most of the observational mass-loss rates
appear within the range of theoretical predictions. Moreover, not only does the
dispersion significantly decrease, but the mean ratio of the observational and
theoretical estimates also becomes closer to one (Table~\ref{peknatabulka}). The usage of
actual observationally determined abundances plays a significant role in this
improvement (Table~\ref{peknatabulka}).

We tested whether the inability of the fitting formula to provide accurate results was caused by a too broad range of input parameters involved in the formula
itself and if the fitting formula restricted to the LMC O stars
provided a better result. We fitted the LMC mass-loss rate predictions of
\citet{mcmfkont} by a simple linear formula $\log\dot M=a+b\log(L)$, with $a$ and
$b$ being free parameters. This provided a similar rms $\chi$ as the more
elaborate fitting formula in Fig.~\ref{hawfit}. This means that the fitting is not improved when the
stellar sample is restricted by metallicity. On the other
hand, when the sample is restricted by luminosity, for instance by $5.4 <
\log(L/L_\odot) < 5.7$, the rms value decreases by more than a half.

The fitting formula of \citet{zcmfkont} does not account for the dependence of the
mass-loss rate on stellar mass, while some other formulae do. However, the
\citet{bjorevol} formula gives a similar rms as the formula of \citet{zcmfkont},
although it includes an explicit dependence on the stellar mass. This indicates
that the higher dispersion of the mass-loss rates from the fitting formula
compared to the dispersion of tailored models is not caused by the omission of
mass from the fits.

\begin{figure}
\includegraphics[width=0.5\textwidth]{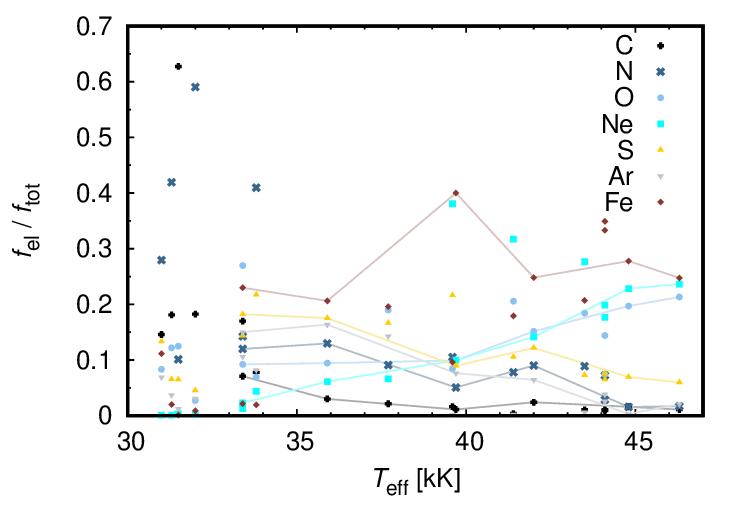}
\caption{Relative contribution of elements that drive the wind most efficiently
to the total line force as a function of the effective temperature. The
contribution was calculated at the critical wind point for all tailored models
from Table~\ref{hvezpar} with actual abundances. The radiative force was
estimated using the Sobolev approximation for computational reasons. The lines
colored corresponding to individual elements connect data points for stars
with a restricted range of luminosities of $5.4 < \log(L/L_\odot) < 5.7$.}
\label{jezero}
\end{figure}

To further understand the origin of the scatter in the mass-loss rate
predictions, we plot the contribution of individual elements to the total line
force in Fig.~\ref{jezero}. The plot shows complex nonmonotonic trends with
temperature, which might contribute to the scatter in the mass-loss rate. A detailed
analysis of the model results shows that the predicted mass-loss rates from the
models with tailored parameters are typically higher than the rates from the
fitting formula for models in which C and Fe contribute more significantly to the radiative force. This indicates that the scatter in the mass-loss rates might be
partially caused by the varying contributions of individual elements to the
radiative force.

When the scatter in the mass-loss rate predictions is attributed to complex variations
in the contribution of individual elements to the radiative force, a natural
question arises about the nature of these variations. Besides the obvious origin
in the change of ionization and excitation state of the matter, the likely
origin of all these variations is the ambiguous dependence of the stellar parameters on
the mass or temperature\footnote{By the ambiguous dependence of stellar parameters
on the mass or temperature, we mean that in many cases, stars from the
\citet{hawvmm} sample appear to lie relatively close to each other in the
Hertzsprung-Russell diagram despite their different masses.}. To test this, we
connect points in Fig.~\ref{jezero} that correspond to a restricted range of
luminosities of $5.4 < \log(L/L_\odot) < 5.7$. The plot now shows mostly monotonic
variations with temperature.

\begin{figure}
\includegraphics[width=0.5\textwidth]{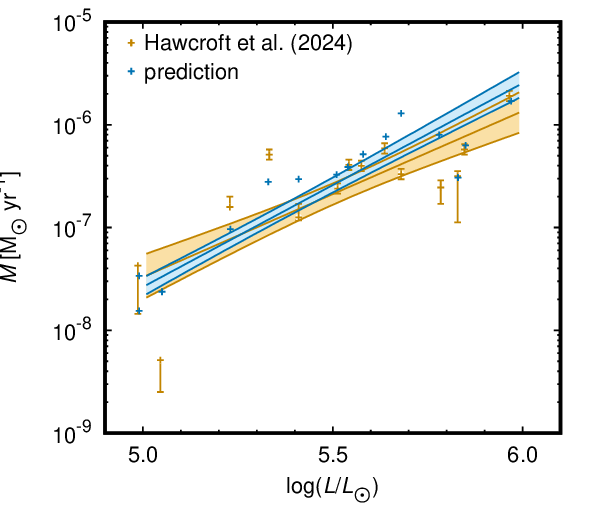}
\caption{Comparison of the predicted and observational dependence of the mass-loss rate
on the luminosity, with their linear fits and corresponding dispersions.}
\label{dmdtlhaw}
\end{figure}

Figure~\ref{dmdtlhaw} compares the predicted and observational dependence of
the mass-loss rate on the luminosity and includes the dispersion of the corresponding relations.
The plots shows significant dispersion of the theoretical relation, which is
comparable to the observational dispersion, but slightly lower. This indicates that
part of the dispersion that appears in the observational dependence of the mass-loss
rate on the luminosity is related to the complex dependence of the mass-loss rate on
the stellar parameters. 

We did not include stars in our analysis so far that feature the weak-wind problem
(they occupy the last four lines in Table~\ref{hvezpar}). Even when we used the
observational stellar parameters, the models predicted mass-loss rates that were higher by
one to two orders of magnitude than inferred from observations.
Theoretical analysis points to the extremely long cooling length in the wind-shock
regions as a possible cause of this discrepancy
\citep{cobecru,nlteiii,predbehli}. As a result of the lower wind density and the less
efficient cooling, the weak-wind problem is expected to appear at higher luminosities
for stars with lower metallicities \citep{zcmfkont}. However, the observational
analysis based on different infrared lines gives contradictory results
\citep{zakonici,mirslabweb}. The origin of this problem therefore remains
open.

\begin{figure}
\includegraphics[width=0.5\textwidth]{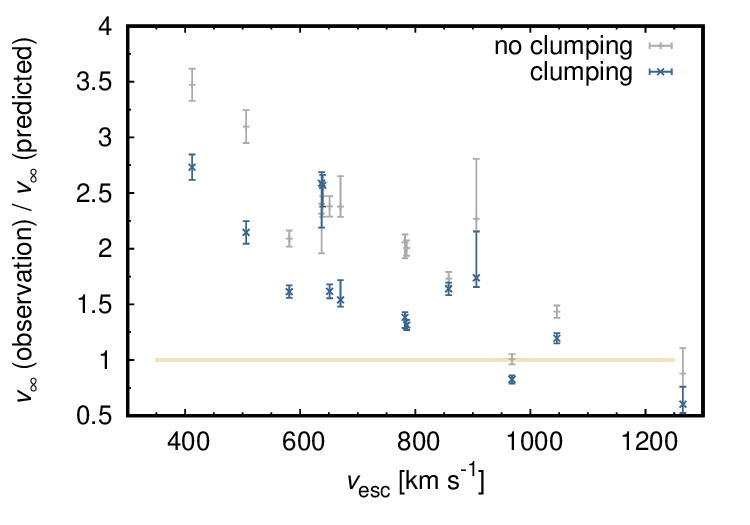}
\caption{Ratio of the observational and predicted terminal velocities for
the stars we studied, plotted for models with and without clumping as a
function of the escape speed. The line denotes the unity ratio.}
\label{vnekhaw}
\end{figure}

Theory predicts that the ratio of the terminal velocity and escape speed should
be basically constant \citep{cak}. Instead, the \citet{hawvmm} sample shows a
decrease in the observational ratio of the terminal velocity and escape speed
with surface gravity or escape speed. This result cannot be easily reconciled
with the theoretical model. The \citet{zcmfkont} models systematically underestimate the
predicted terminal velocities. The use of the appropriate stellar parameters did not
lead to a systematic shift in the terminal velocities, and consequently, the problem
that the predicted terminal velocities are too low remains even in tailored models
(Fig.~\ref{vnekhaw}). The discrepancy is stronger for stars with a weaker surface
gravity. To explain this effect, the effect
of X-rays or radially variable clumping might be considered \citep[e.g.,][]{lojza,irchuch,savi}.

\section{The effect of clumping}

\begin{table}
\caption{Predicted wind mass-loss rates and terminal velocities determined from
smooth models (denoted without any subscript) and from models with clumping
according to the observational constraints (denoted with the subscript
cl).}
\centering
\label{hvezparch}
\begin{tabular}{rcccc}
\hline
\hline
ID & $\dot M$ & \vnek & $\dot M_\ch$ &  $\varv_{\infty,\ch}$\\\
& [\msr] & [\kms] & [\msr] & [\kms] \\
\hline
087 & $9.7\times10^{-8}$ & 1390 & $9.6\times10^{-8}$ & 1800 \\
440 & $7.7\times10^{-7}$ &  690 & $1.1\times10^{-6}$ &  880\\
223 & $1.5\times10^{-8}$ &  930 & $1.5\times10^{-8}$ & 1210\\
664 & $3.9\times10^{-7}$ & 1120 & $4.8\times10^{-7}$ & 1000\\
385 & $2.8\times10^{-7}$ & 1090 & $3.7\times10^{-7}$ & 1050 \\
184 & $2.4\times10^{-8}$ & 2180 & $2.5\times10^{-8}$ & 2670 \\
244 & $5.2\times10^{-7}$ & 1090 & $4.9\times10^{-7}$ & 1610 \\
096 & $3.1\times10^{-7}$ & 2620 & $3.6\times10^{-7}$ & 3820 \\
422 & $1.3\times10^{-6}$ &  680 & $1.4\times10^{-6}$ &  980 \\
608 & $6.3\times10^{-7}$ & 1810 & $6.4\times10^{-7}$ & 2170 \\
180 & $1.7\times10^{-6}$ & 1090 & $1.7\times10^{-6}$ & 1690 \\
216 & $8.0\times10^{-7}$ & 1410 & $8.8\times10^{-7}$ & 2100 \\
143 & $3.3\times10^{-7}$ & 1680 & $3.4\times10^{-7}$ & 1770\\
586 & $3.0\times10^{-7}$ & 1450 & $3.1\times10^{-7}$ & 2210\\
\hline
517 & $3.4\times10^{-8}$ & 1550 & $3.2\times10^{-8}$ & 2550 \\
235 & $9.3\times10^{-10}$& 3810 & $1.1\times10^{-9}$ & 4120 \\
627 & $8.0\times10^{-9}$ & 1510 & $1.0\times10^{-8}$ & 3760 \\
280 & $1.4\times10^{-8}$ & 1570 & $1.3\times10^{-8}$ & 2730 \\
\hline
\end{tabular}
\tablefoot{As in Table~\ref{hvezpar}, the horizontal line separates stars with
normal mass-loss rates from stars that display the weak-wind problem.}
\end{table}

In addition to the wind mass-loss rates and terminal velocities, \citet{hawvmm}
also determined the radial variations of the clumping parameters for individual stars. We used this information to infer the effect of clumping on
the predicted wind properties, and we tested in particular whether the radial variations
in the clumping can reproduce the higher observed terminal velocities.

We introduced the radially varying clumping parameter
$\cc={\langle\rho^2\rangle}/{\langle\rho\rangle^2}$ into our models. Following
\citet{hawvmm}, we assumed a smooth wind $\cc=1$ up to the wind velocity
$\vel_\ch$, from which the clumping parameter starts to grow linearly and
achieves a maximum value at $2\vel_\ch$. From this point, the clumping parameter
is again constant and equal to its maximum value. The corresponding parameters are
given in Table~2 of \citet{hawvmm}. To avoid problems with the convergence, we
fitted the model radial velocity of smooth models by \citep{betyna,bcmfkont}
$\tilde \vel (r)=\sum_{i=1}^2 \varv_i\zav{1-\gamma{R_*}/{r}}^i$, where $\varv_i$
and $\gamma$ are free parameters of the fit. We adopted the $\vel_\ch$ value from
\citet{hawvmm}, expressed relative to the wind terminal velocity, for which we
used the value from the fit of the corresponding wind model,
$\tilde\varv_\infty=\varv_1+\varv_2$.

To study the wind clumping, we restricted ourselves only to the models with
light element abundances derived from spectroscopy. The resulting wind parameters
derived from the models that include clumping are compared to smooth models in
Table~\ref{hvezparch}. Because the clumping is assumed to start at high wind
velocities, its effect on the mass-loss rate is low. On the other hand,
clumping shifts the ionization balance toward ions with a lower charge, which
accelerate the wind more efficiently. As a result, the wind terminal velocity
increases toward the values inferred from observations (Fig.~\ref{vnekhaw}).

\section{Conclusions}

We calculated detailed line-driven wind models with parameters corresponding to the
LMC stars studied by \citet{hawvmm}. Our aim was to understand whether the scatter
that appears in the relation between the observational values of the wind
mass-loss rate and the luminosity is real or just a result of the uncertainties
connected with a spectral analysis.

The tailored wind models parameterized by observational values reproduce the
observations better than the fitting relations. This indicates that part of the
scatter seen in the observational data could be real and can be attributed
to the star-to-star variations in the effective temperature, radius, and
metallicity. These variations lead to nonmonotonic variations in the ionization
fractions and in the contribution of individual elements to the radiative force. This
sensitivity of the radiative force to the ionization fractions contributes to the
dispersion of the relation between the mass-loss rate and luminosity. However, a
part of the dispersion still remains unaccounted for.  At least part of this
remaining dispersion might perhaps be connected to the time variability of the wind
line profiles \citep{masvar,parvar}. Another possibility is that the
observational analysis is based on different atmosphere codes featuring various
assumptions, which may add to the scatter of the mass-loss rates inferred from
observations \citep{sanporov}.

On the other hand, even the models that were computed with observational stellar parameters
underestimate the terminal velocity with respect to the observational values.
This disagreement can be partially alleviated in models with radially dependent
clumping.

\section*{Data availability}

Wind and atmosphere model output files are available in electronic form at the CDS via
anonymous ftp to cdsarc.u-strasbg.fr (130.79.128.5) or via
http://cdsweb.u-strasbg.fr/cgi-bin/qcat?J/A+A/.

\begin{acknowledgements}
This work was supported by grant GA \v{C}R 25-15910S. Computational resources
were provided by the e-INFRA CZ project (ID:90254), supported by the Ministry of
Education, Youth and Sports of the Czech Republic. 
\end{acknowledgements}

\bibliographystyle{aa}
\bibliography{papers}

\end{document}